**Title: Effect of COVID-19 on noise pollution change in Dublin, Ireland**

Bidroha Basu[1,2,*], Enda Murphy[1], Anna Molter[1,3], Arunima Sarkar Basu[1], Srikanta Sannigrahi[1], Miguel Belmonte[4], Francesco Pilla[1]

[1]School of Architecture, Planning and Environmental Policy; University College Dublin, Ireland.

[2]Department of Civil, Structural and Environmental Engineering; Trinity College Dublin, Ireland.

[3]School of Environment, Education and Developmemt; The University of Manchester, Manchester, UK.

[4]School of Electrical and Electronic Engineering; The University of Manchester, Manchester, UK.

**Abstract**

Noise pollution is considered to be the third most hazardous pollution after air and water pollution by the World Health Organization (WHO). Short as well as long-term exposure to noise pollution has several adverse effects on humans, ranging from psychiatric disorders such as anxiety and depression, hypertension, hormonal dysfunction, and blood pressure rise leading to cardiovascular disease. One of the major sources of noise pollution is road traffic. The WHO reports that around 40% of Europe's population are currently exposed to high noise levels. This study investigates noise pollution in Dublin, Ireland before and after the lockdown imposed as a result of the COVID-19 pandemic. The analysis was performed using 2020 hourly data from 12 noise monitoring stations. More than 80% of stations recorded high noise levels for more that 60% of the time before the lockdown in Dublin. However, a significant reduction in average and minimum noise levels was observed at all stations during the lockdown period and this can be attributed to reductions in both road and air traffic movements.

# 1. Introduction

The growth in urban populations around the world has resulted in the expansion of cities which are important for providings jobs, housing, and sustainable livelihoods (He et al., 2018; Forman and Wu, 2016; Zhou et al., 2016; Sun et al., 2016; Brueckner, 2000; Peng, 1997). Given the evolution of cities and particularly those that are more heavily dependant on car-based travel, pollution levels within cities has shown an increasing trend over time (Hien et al., 2020; Lagonigro et al., 2018; Rocha et al., 2017). Amongst all pollutants, the World Health Organisation (WHO) has reported that noise pollution is the third most hazardous type of pollutant after air and water pollution (WHO, 2005). Health impacts due to an increase in environmental noise are a concern worldwide (Alves et al., 2015; Merchan et al., 2014; Nedic et al., 2014; Ongel and Sezgin, 2016; Pathak et al., 2008). For example, noise sensitivity can be an important contributor to psychiatric disorders such as anxiety and depression (Belojević et al., 1997; Fyhri and Aasvang, 2010; Ongel and Sezgin, 2016). Recent studies suggest that an increase of 5dB roadside noise can raise the chance of hypertension by 3.4% (Kim et al., 2019; Oh et al., 2019; Eriksson et al., 2012). Further studies indicate that exposure to a high level of noise can result in hormonal dysfunction and can also contribute to the rise in blood pressure which can severely impact the cardiovascular system in the body (Said and Goharty, 2016; Münzel and Sørensen, 2017). Other studies have reported that pregnant women might be at a greater risk of being affected to noises since they are more sensitive to environmental stress factors (Selander et al., 2019; He et al., 2019; Poulsen et al., 2018; Sears et al., 2018; Murphy and Faulkner, 2018). Ashin et al. (2018) noted that road traffic noise can increase gestational diabetes mellitus, which leads to glucose intolerance that occurrs during the beginning of pregnancy. Based on 29 case study analysis where pregnant women were exposed to 80dB or higher noise, Dzhambov et al. (2014) documented that the risk for having gestational hypertension, small for gestational age and babies with congenital malformations increases significantly. The European Union documented that more than 40 percent of the total European population is exposed to a Day-Evening-Night ($L_{den}$) noise level of 55dB or greater, while 30 percent of the population is exposed to the same noise level during night-time (WHO, 2017; Foraster et al., 2017; Pitchika et al., 2017; Basner and Mcguire, 2018). A city soundscape study (Lebiedowska, 2005; Léobon, 1995) suggested that urban noise can be broadly classified in the following four categories: i) background noise: classified as unpleasant due to the presence of high pitched, piercing, strong, continuous, irregular, or intermittent noises that cause humming of the peripheral environment; ii) mechanical noise: noise caused by mechanical equipments such as vehicles, railway, and aircraft as well as large industrial plants generating noises; iii) human activity related noise: can generate from demonstrations, gatherings, sirens, trades, household noise due to usage of vacuum cleaners or drills etc.; iv) other environmental noise: can be attributed to the presence of storms, thunders, winds, and creaking.

The Noise Observation and Information Service for Europe (NOISE) suggest that the majority of the noise affecting the exposure of the population are being generated by road vehicle traffic, compared to other sources of transportation (King and Murphy, 2016; Śliwińska-Kowalska and Zaborowski, 2017; Nedic et al., 2014; Voltes and Martin, 2016; Yuan et al., 2019; Das et al., 2019). Detailed studies and literature reviews have shown that the noise pollution from urban road traffic has the highest level of exposure as the roads are located in close proximity to built infrastructures such as schools, offices and residential buildings (Khan et al., 2018; Paiva et al., 2019; von Graevenitz, 2018; Rey and Barrigón, 2016; Cai et al., 2019; González et al., 2018; Ruiz-Padillo et al., 2016; Zannin et al., 2003). Moreover, research has shown that not only the noise pollution, but the duration of exposure to traffic-related noises has impacted the health conditions of people severely (Tonne et al., 2016; Buxton et al., 2017; Khan et al., 2018; Cai et al., 2019). The European Commission has reported that the noise from rail and road costs the European Union around 40 billion euros per year in terms of socio-economic damages (Vaitkus et al., 2016; Andriejauskas et al., 2018). Apart from socio-economic cost, studies have also revealed that noise pollution caused due to human-generated factors can potentially alter the biodiversity by impacting the distribution and behavior of species and their habitat quality (Siemers and Schaub, 2011; Estabrook et al., 2016; Guttal and Jayaprakash, 2007). Although noise levels are generally highest around the roadways and near the transportation terminals, the exposure of the large majority of city inhabitants is determined by a city's overall background noise (Mehdi et al., 2011; Stansfeld and Matheson, 2003; Khan et al., 2018; Raimbault et al., 2003; Morillas et al., 2002). Other studies have further revealed that several factors such as morphology of the city (aspect ratios of buildings), current and projected population as well as household density, noise regulations laws and policies, traffic network design, availability of affordable public transport nearby of the area, frequency of traffic jams, the public vehicles to private vehicles on road, the average insulation of the homes and the noises escaping from the building, and the motorized driving behavior, possibility of construction of new buildings, etc. can determine the noise levels at different scales (Ariza-Villaverde et al., 2014; Bouzir and Zemmouri, 2017; Piccolo et al., 2005; Abbaspour et al., 2015). Noise distribution assessment in a city should also take into the local behavior of traffic flows, street, and urban maintenance or expansion initiatives, if being undertaken (Mato and Mufuruki, 1999; Yu and Kim, 2011; Barros and Dieke, 2008). Thus, noise mapping studies need to be conducted at the microscale as well as at a broader scale (Margaritis and Kang, 2016; Dutta et al., 2017; Tiwari et al., 2019). Other studies reveal that the material building blocks used for the construction of building infrastructure as well as building and street design at the nearby vicinity and neighborhood quality are correlative factors responsible for determining the level of noise (Onaga and Rindel, 2007; Picaut and Simon, 2001; Salomons and Pont, 2012). Similar studies suggest that the closed building blocks with low population density lead to lower noise levels than open building block constructions with higher population density (Silva et al., 2018; Zhou et al., 2017; de Souza et al., 2011; Costa et al., 2011; Nega et al., 2012). The percentage of quiet areas is correlated to the nature of the street and the influx of traffic and population present in the area (Liu et al., 2019; Shepherd et al., 2013).

Within the foregoing context, this study investigates the effect of the COVID-19 lockdown on changes in noise pollution in Dublin, Ireland. Since the lockdown, the traffic volume has been substantially reduced and social events have been restricted considerably; therefore, a reduction in noise pollution across the city would be expected. This study performs a regression based trend analysis to investigate the changes in noise pollution before and after the lockdown. Furthermore, a change-point analysis was performed to identify the exact date on which the change in the pattern of noise pollution occurred in the noise monitoring stations; change-point analysis is useful for estimating the point at which the statistical properties of a sequence of observations change.

## 2. Methodology

An analysis of changes in noise pollution in Dublin, Ireland due to the lockdown is performed based on noise data obtained from 12 monitoring stations in Dublin. This section provides a detailed description of the data and the methodology used in the analysis.

### 2.1. Data details

Equivalent Continuous Sound Pressure Level ($L_{eq}$) and maximum noise ($L_{max}$) data recorded at 5 minute intervals at 12 locations in Dublin (Figure 1 and Table 1) were collected from 1[st] January 2020 until 11[th] May 2020 in decibel (dB). The $L_{eq}$ is the average noise rrecorded within the 5 minute interval a nd the $L_{max}$ is the maximum noise recorded in the time interval. An hourly maximum ($L_{max,1hr}$), hourly minimum ($L_{min,1hr}$)and hourly average noise ($L_{eq,1hr}$) was extracted for each of the 12 stations based on the $L_{eq}$ and maximum noise data. The hourly maximum noise for the chosen hour (in dB) was obtained from the 5 min maximum time scale data by considering the highest noise recorded during the selected hour period. The minimum hourly noise data were extracted based on the lowest recorded $L_{eq}$ noise over an hour's timeframe. The hourly average noise for the selected hour was obtained from $L_{eq}$ based on the following equation (Murphy and King, 2014):

$$Avg - noise = 10 \times log_{10}\left(\frac{\sum_{i=1}^{n} 10^{N_i/10}}{nh}\right)$$

(1)

where $\{N_i, \ i = 1, \cdots, nh\}$ is the $nh$ number of $L_{eq}$ noises during the chosen hour at a 5 min interval. Equation (1) ensures that the hourly average noise produces the same amount of total energy produced over the one-hour period.

Relating high noise with traffic variable is a challenging task. The major difficulty lies in acquiring reliable traffic variable data for Dublin at an hourly scale. In this study, annual average hourly traffic flow data at every road in Dublin were obtained from the Dublin City Council's Sydney Coordinated Adaptive Traffic System (SCATS) for the year 2016. It needs to be noted that 2016 is the latest available traffic data for Dublin, which can be assumed to be a representative traffic flow for the city before lockdown, while considerably lower traffic flow can be assumed during the lockdown period.

## 2.2. Statistical analysis

In order to estimate the changes in noise pollution over the entire time period , two approaches have been used in this study. A linear regression is fitted to estimate the slope of the noise pollution time series. Considering there are $n$ number of hourly noise data points $Y_{[n \times 1]}: \{y_t; \ t = 1, \cdots, n\}$ where $\{h_t = t; \ t = 1, \cdots, n\}$ is the $t$-th hour for a station, and defining

$H_{[n \times 2]} = \begin{bmatrix} 1 & h_1 \\ 1 & h_2 \\ \vdots & \vdots \\ 1 & h_n \end{bmatrix}$ the regression coefficient $\beta = [\beta_0 \quad \beta_1]$ can be estimated as

$$\beta = (H'H)^{-1} H'Y \tag{2}$$

where $\beta_0$ is the intercept and $\beta_1$ is the slope of the regression. A positive value of $\beta_1$ indicates an increase in noise over time, while a negative value denotes a decrease in noise. The statistical significance of the slope is denoted by the p-value associated with the corresponding hypothesis test.

The regression coefficient provides information only on the linear trend of the change in noise over time and does not detect the presence of any sudden change in the noise data. As the lockdown might change the noise pollution pattern within a very short period of time, it is necessary to perform a change-point detection analysis to capture those forms of abrupt change that might be present in the noise data. A change point detection algorithm with a linear computational cost adopted from Killick et al. (2012) was used in this study. Details on the algorithm are as follows.

Let the time series data be denoted as $\{y_t; \ t = 1, \cdots, n\}$ and the change point is $\tau, 1 < \tau < n$, which subdivide the time series into two components $\{y_1, \cdots, y_\tau\}$ and $\{y_{\tau+1}, \cdots, y_n\}$. Identification of $\tau$ can be performed by minimization of the following objective function.

$$C\{y_1, \cdots, y_\tau\} + C\{y_{\tau+1}, \cdots, y_n\} + \gamma \tag{3}$$

where $C$ is a cost function and $\gamma$ is a penalty term to stop overfitting.

It needs to be noted that in situations where $C\{y_1, \cdots, y_\tau\} + C\{y_{\tau+1}, \cdots, y_n\} + \gamma \geq C\{y_1, \cdots, y_n\}$, there is no sudden change in the time series.

The penalty term considered in the study is the Schwarz/Bayesian information criteria and is given by,

$$\gamma = d \times \log(n) \tag{4}$$

where $d$ is the number of additionally introduced model parameters for estimating the change point.

The cost function is provided as,

$$C\{y_i, \cdots, y_j\} = -\max_\theta \sum_{t=i}^{j} \log f(y_t|\theta) \tag{5}$$

where $f(y_t|\theta)$ is the density function of the time series data $\{y_i, \cdots, y_j\}$ and $\theta$ is the set of distribution parameters.

A linear model (linear regression) can be used to model the properties of a time series in situations where the underlying system is linear in nature. A system can be assumed to be linear if the response function $y(t)$ from the system follows the following criteria (Billings and Voon, 1983):

Considering $y'(t) = y(t) - \overline{y(t)}$, if the cross-correlation function between $y'(t)$ and $y'(t)^2$ for any lag $\tau$ are zero (equation 6), the system is assumed to be linear.

$$\phi_{y',y'^2}(\tau) = \frac{\sum_{t=1}^{N-\tau}\left[y'(t) - \overline{y'(t)}\right] \times \left[\left(y'(t+\tau)\right)^2 - \overline{\left(y'(t)\right)^2}\right]}{\sqrt{\sum_{t=1}^{N}\left[y'(t) - \overline{y'(t)}\right]} \times \sqrt{\sum_{t=1}^{N}\left[\left(y'(t)\right)^2 - \overline{\left(y'(t)\right)^2}\right]}}$$

$$= \begin{cases} 0, \forall \tau = 0,1,2,\cdots & \Rightarrow linear\ system \\ \neq 0, \forall \tau = 0,1,2,\cdots & \Rightarrow nonlinear\ system \end{cases}$$

$$(6)$$

For large data (considerable high value of $n \geq 30$), the 95% confidence interval can be chosen as $\pm\ 1.96/\sqrt{n}$, indicating values of $\phi_{y',y'^2}$ within this bound can be assumed to follow a linear system.

## 3. Results

Hourly average noise obtained from the 12 noise monitoring stations for the pre-lockdown period ranging from 1st January 2020 to 24th March 2020 was compared to the hourly average noise during lockdown from 25th March 2020 (Figure 2). Figure 2 indicates that the average noise level reduced once the lockdown started for each of the 12 stations. The noise was found to be higher at at Ballymun Library (ID #2), Chancery Park (ID #4) and Dolphin's Barn (ID #6) before the beginning of the lockdown, while stations at Drumcondra Library (ID #7) and Woodstock Gardens (ID #12) has the lowest noise pollution levels. Reduction in the average noise during the lockdown period (25th March 2020 to 11th May 2020) are highest at DCC Rowing Club (ID #5), Navan Road (ID #9) and Woodstock Gardens (ID #12), while the lowest reductions occurred at Mellows Park (ID #8) and Raheny Library (ID #10). To investigate the noise reduction further, two boxplots were prepared for each station representing the pre-lockdown (case A) and lockdown (case B) periods. The boxplots for each station were prepared separately based on the hourly average, hourly maximum, and hourly minimum noise data (Figure 3). Comparison of the boxplots indicates that the median (shown in red), 25th and 75th percentile (lower and upper value of the boxes) of noise pollution are lower for all stations and for average, maximum as well as minimum noise pollution. The $5^{th}$ and $95^{th}$ percentile (extreme ends of the whiskers) are lower for average and maximum noise pollution for all stations except for Blessington Street Basin (ID #3) which has a slightly higher $95^{th}$ percentile quantile for maximum noise.

Figure 4 subdivides the time series of the noise data into two components based on the results from the change point detection analysis. The vertical green line indicates the date and time on which a sharp change in the noise occurred. It can be noted from the figure the local average noise level (shown in red dot linear trend) at the period on which the change point occurred, reduced considerably for all of the stations. The date for which the sudden reduction in noise happened is provided in Table 1. It indicates that 7 stations ( IDs #3, 4, 5, 6, 7, 11, 12) had a sudden noise reduction between 14-17$^{th}$ March 2020, while the other 5 stations (IDs #1, 2, 8, 9, 10) had a sharp reduction in noise closer to the lockdown date. It can be noted that the stations exhibiting early reduction in noise are located inside the city and closer to the city center (except Walkinstown Library, ID #11), while the other 5 stations exhibiting noise reduction closer to the lockdown period are located near the boundary of the city and close to the M50 motorway. Since the noise pattern varies considerably on the time of the day due to changes in traffic conditions and other human activities, the 24-hour period is subdivided into three timeframes, night time (11 PM to 7 AM), day time (7 AM to 7 PM) and evening time (7 PM to 11 PM). Average, maximum, and minimum noise for each of those three timeframes for each station were estimated based on the hourly average noise data. Subsequently, a regression analysis was performed for each station corresponding to every timeframe to identify the trend in changes in the noise. Figure 5 shows the time series plots (shown in black) and the trend (shown in red), of the average noise, for brevity. The regression slope values for average, maximum, and minimum noise for each station and for every timeframe are provided in Table 2. A negative slope indicates a decrease in noise pollution over time, while positive value denoted that the noise increases during the lockdown period. Slopes that are statistically significant at 95% confidence interval (5% significance level in a two-tail test) are shown in bold, while the remaining slope values (both positive and negative) can be statistically considered to be equal to zero, and hence are discarded from further analysis. A linear regression can be used to understand the changes in noise pollution over time in situations where the underlying system generating the noise pollution can be considered to be linear in nature. The cross-correlation function $\phi_{y',y'^2}(\tau)$ corresponding to lags ranging from 1 to 50 for each station corresponding to night, day and evening time was plotted (Figure 6). The figures indicated that majority of the time series data can be considered to be linear in nature.

Traffic is considered to be one of the major sources of noise in a city. The average traffic volume at each street in Dublin for night, day, and evening time is shown in Figure 7. Average traffic counts surrounding each noise monitoring station within a radius of 500 meters for night time, day time and evening time were estimated (Table 3) and compared to the hourly average noise pollution in the pre-lockdown period using a scatterplot (Figure 8). The figure indicates that except Ballymun Library (ID # 2) and DCC Rowing Club (ID #5), noise in all the other stations can be considered to be related to the traffic volume. The figure can be subdivided into two components, where the four stations Blessington Street Basin (ID #3), Chancery Park (ID #4), Drumcondra Library (ID #7) and Woodstock Gardens (ID #12) that are located closer to the city center and exhibit hight traffic flow can be grouped together in relation with traffic flow, while the remaining 6 stations (Ballyfermot Civic Office, ID #1; Dolphin's Barn, ID #6; Mellows Park, ID #8; Navan Road ID, #9; Raheny Library, ID #10; Walkinstown Library, ID #11) with relatively lower traffic flow follows a similar pattern. A linear regression line and the corresponding $R^2$ value are being added in the figure for brevity. The $R^2$ value indicates that the noise in the four stations (ID #3, 4, 7, 12) located near the city center is highly related to the traffic, while the other 6 stations is somewhat dependend on traffic. It should be noted that Ballymun Library (ID # 2) is closer to the Dublin airport and DCC Rowing Club (ID #5) is located adjacent to the river Liffey, hence noise sources such as aeroplanes and waterbodies, respectively, can have a major impact on the noise pollution in those two stations.

According to WHO guidelines (WHO, 2009), any noise above a threshold value of 55 dB is considered to be harmful for the human population. Long term exposure to noise level above the threshold value leads to frequent occurrence of health adversity to a large part of the exposed population. The total number and the percentage of times the hourly average noise pollution exceeded the threshold 55dB for every monitoring station before and during lockdown is shown in Table 4. The table indicated that except Drumcondra Library (ID #7) and Woodstock Gardens (ID #12), every monitoring stations had recorded noise pollution greater than the threshold for more than 60% of the time, while for Ballymun Library (ID #2) and Chancery Park (ID #4), more than 95% of the time the hourly average noise exceeded the threshold value. During lockdown, the number and the percentage of times the noise exceeded the threshold reduced considerably for each monitoring station. During the lockdown period, hourly average noise pollution at only 3 stations (Ballymun Library, ID #2; Chancery Park, ID #4; Dolphin's Barn, ID #6) exceeded the threshold for more than 60% of the time.

## 4. Discussion

A higher value of noise pollution in Ballymun Library (ID #2) can be attributed to the fact that the monitoring station is located in close proximity to Dublin airport and the M50 highway, which is the busiest street in Dublin (Figure 1). The other two locations (Chancery Park, ID #4 and Dolphin's Barn, ID #6) exhibiting higher noise pollution are located close to the city center area. Furthermore, Chancery Park is adjacent to the Liffey river, where waterbody contributes to noise pollution. On the other hand, Drumcondra Library (ID #7) and Woodstock Gardens (ID #12) are located away from major roads (Figure 7) and have the lowest noise pollution. During the lockdown, DCC Rowing Club (ID #5), Navan Road (ID #9) and Woodstock Gardens (ID #12) had the highest reduction in noise (Figure 2 and Figure 3a), which can be attributed to the fact that DCC Rowing Club was shut down during the lockdown period, and the major source of noise in the station is the water flow in the river Liffey. Navan Road is located adjacent to the Phoenix Park, which is the biggest park in Dublin and was closed. Also, people living in the Lucan area located in the northwest part of the city mainly use this road to travel to the city center for jobs and other activities. Woodstock Gardens, on the other hand, is located in close proximity to University College Dublin that has the biggest university campus in Dublin, which was closed during the lockdown. Reduction in noise at Mellows Park (ID #8) and Walkinstown Library (ID #11) were found to be least (Figure 3a) since those monitoring stations are in close proximity to the M50 highway.

Monitoring stations at Blessington Street Basin (ID #3), Chancery Park (ID #4), DCC Rowing Club (ID #5), Dolphin's Barn (ID #6), Drumcondra Library (ID #7), Walkinstown Library (ID #11) and Woodstock Gardens (ID #12) had a sudden decrease in noise pollution between 14-17[th] March 2020 (Figure 4 and Table 1). For precautionary measures, the majority of the primary and secondary schools in Dublin were closed from 14[th] March 2020. The number of schools within 500 meter radius distance from each monitoring station are shown in Table 1. Except DCC Rowing Club (ID #5) and Walkinstown Library (ID #11), all the other stations are surrounded by several primary and secondary schools (Table 1), while DCC Rowing Club and Walkinstown Library was closed by mid-March for precautionary measure. One point to note is that even though Ballyfermot Civic Office (ID #1) and Mellows Park (ID #8) monitoring stations are also located close to several schools, Ballyfermot Civic Office is located close to the M50 (Figure 1), while Mellows Park is close to an industrial area in Dublin and the M50 highway. Those two stations along with Ballymun Library (ID #2), Navan Road (ID #9) and Raheny Library (ID #10) had a reduction in noise noted between 19-29[th] March 2020, when the lockdown process slowly started, where all those stations are located close to the M50.

Statistically significant slope values for average noise and minimum noise for night time, day time, and evening time indicate that the average and minimum noise reduced for each monitoring station due to the lockdown. However, an increase in maximum noise pollution during the night time was noted at Ballyfermot Civic Office (ID #1) and DCC Rowing Club (ID #5). Both the stations are located in the western part of the city and close to the Phoenix park. One reason might be an increase in wildlife activities during night time in the park after lockdown.

Since the traffic information is unavailable for Dublin during the lockdown period, the traffic data was linked with noise pollution for the pre-lockdown period. Furthermore, since the traffic data was not available for 2020, the best available data were used as an approximation. As the Ballymun Library (ID #2) is close to the M50 but it is more than 500 meters in distance (~950 meters), the traffic count does not get reflected. Also, one of the major noise sources for this station is the flights since it is in close proximity to the Dublin airport, which can be the reason for different behavior of the station compared to the other stations. However, even though the scatterplot denotes the presence of a relation between noise and traffic for all the other stations, it needs to be noted that traffic alone cannot be considered to be the sole factor generating noise, and other factors need to be explored to model noise pollution, which is evident from the amount of variability present in those scatterplots (Figure 8). The correlation between traffic count and noise pollution is considerably high only for the four stations located surrounding the city center (ID #3, 4, 7, 12).

As Drumcondra Library (ID #7) and Woodstock Gardens (ID #12) are located in regions with low traffic, the number of times the noise position exceeded the threshold of 55 dB were low, even before the lockdown. Since Ballymun Library (ID #2) is located close to the airport and M50, and Chancery Park (ID #4) is located in the city center, both the locations receive considerably high noise pollution, which reduced from 97-98% to 82-83% during lockdown. Ballyfermot Civic Office (ID #1), Rowing Club (ID #5) and Navan Road (ID #9) had the highest reduction of high hourly average noise (47-63%) during locadown. It can be noted that those three stations are located in the west Dublin and close to the Phoenix park.

Based on the analysis it can be noted that the lockdown had indeed reduced the noise pollution considerably in all the 12 monitoring stations. The relation between noise pollution and variables such as traffic volume, location of river, time of the day etc. are related to each other, where they have an impact in the noise pollution.

## 5. Conclusion

This study investigated the effect of the COVID-19 lockdown on noise pollution in Dublin, Ireland. The results show a reduction in noise pollution at all 12 noise monitoring stations in Dublin. It is likely that this reduction in noise is due to a decrease in road and air traffic during the lockdown; however, these are not the only factors determing noise levels, and noise due to factors such as weather and natural phenoma do impact measurements at monitoring stations. This study has several important implications: it demonstrates that a significant ($p<0.05$) reduction in average and minimum noise can be achieved in a major urban area through a reduction in traffic. Although it may not be feasible to completely eliminate traffic from cities, it shows that alternative modes of transport, such as walking and cycling, should be promoted to improve sustainability of the transport infrastructure. It is well established that noise pollution is associated with ill health; therefore, city-wide reductions in noise could provide important public health benefits. Further analyses of potential health impacts associated with the reduction in noise in Dublin will be carried out in the future. Finally, this study demonstrates the value of a city-wide permanent noise monitoring network and suggests that noise monitoring networks should be implemented in urban areas around the world.

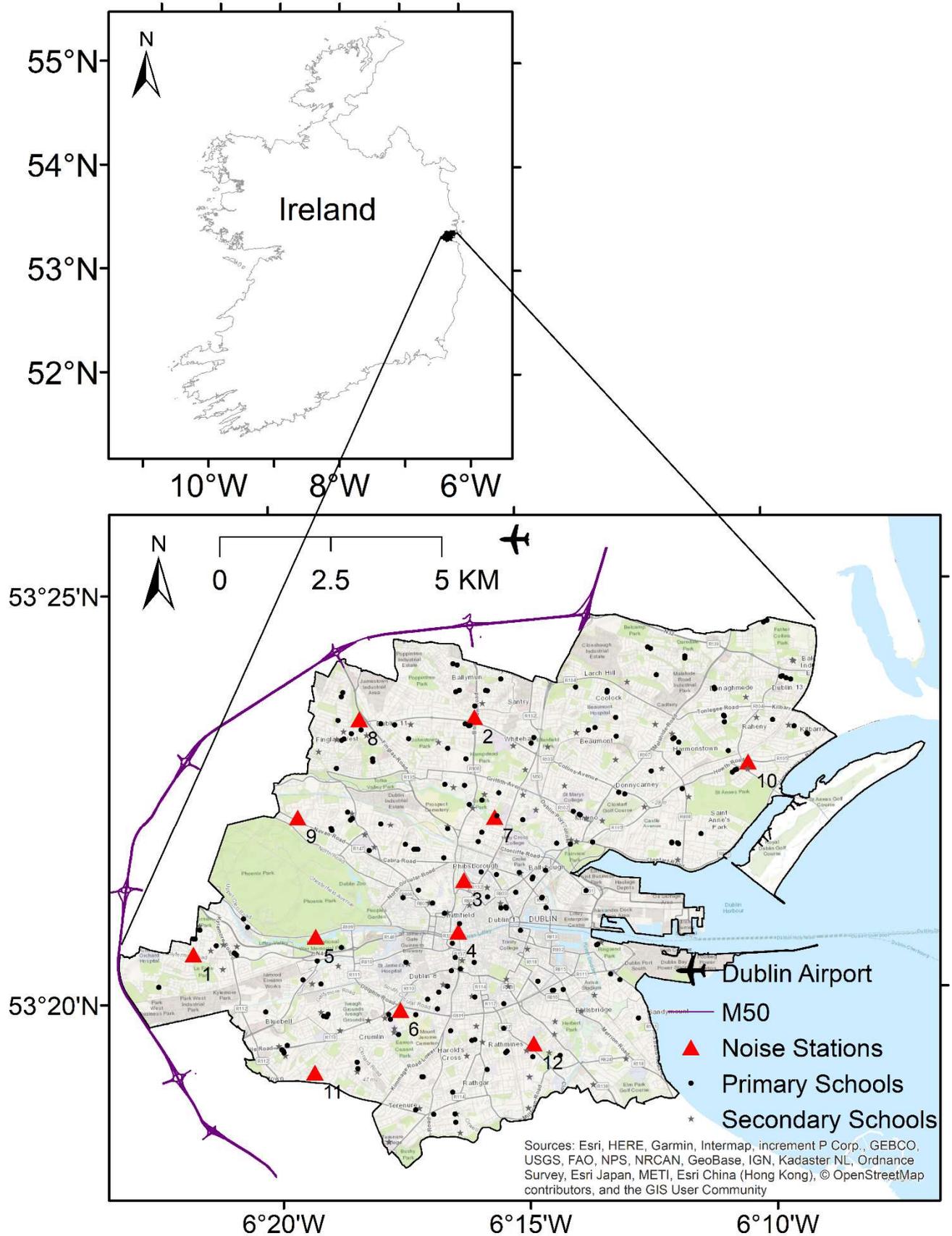

**Figure 1.** Locations of the noise monitoring stations and the primary and secondary schools at Dublin, Ireland.

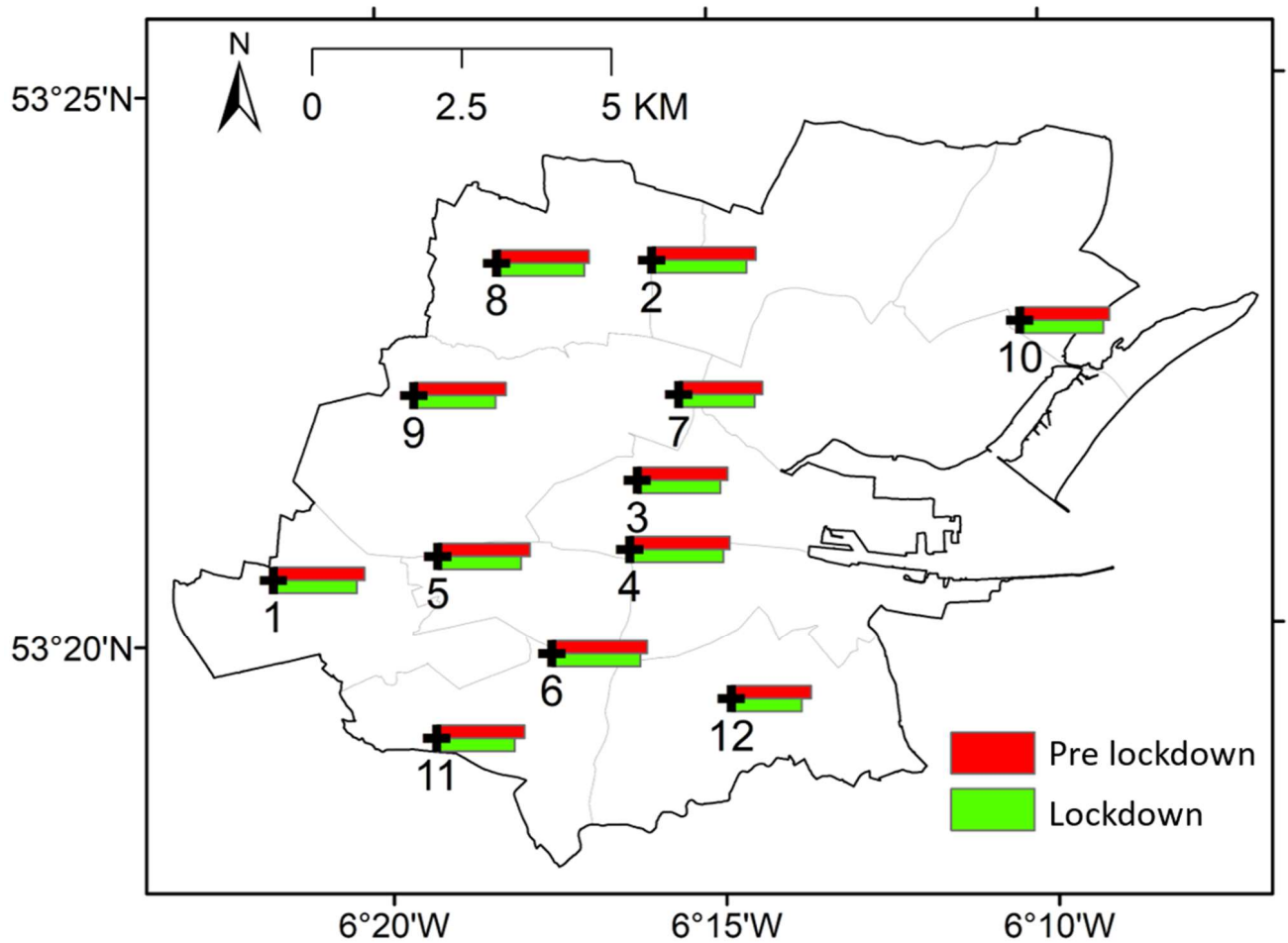

**Figure 2.** Average noise pollution at monitoring stations in Dublin for pre lockdown period (1st January – 24th March 2020) and during lockdown (25th March – 11th May 2020).

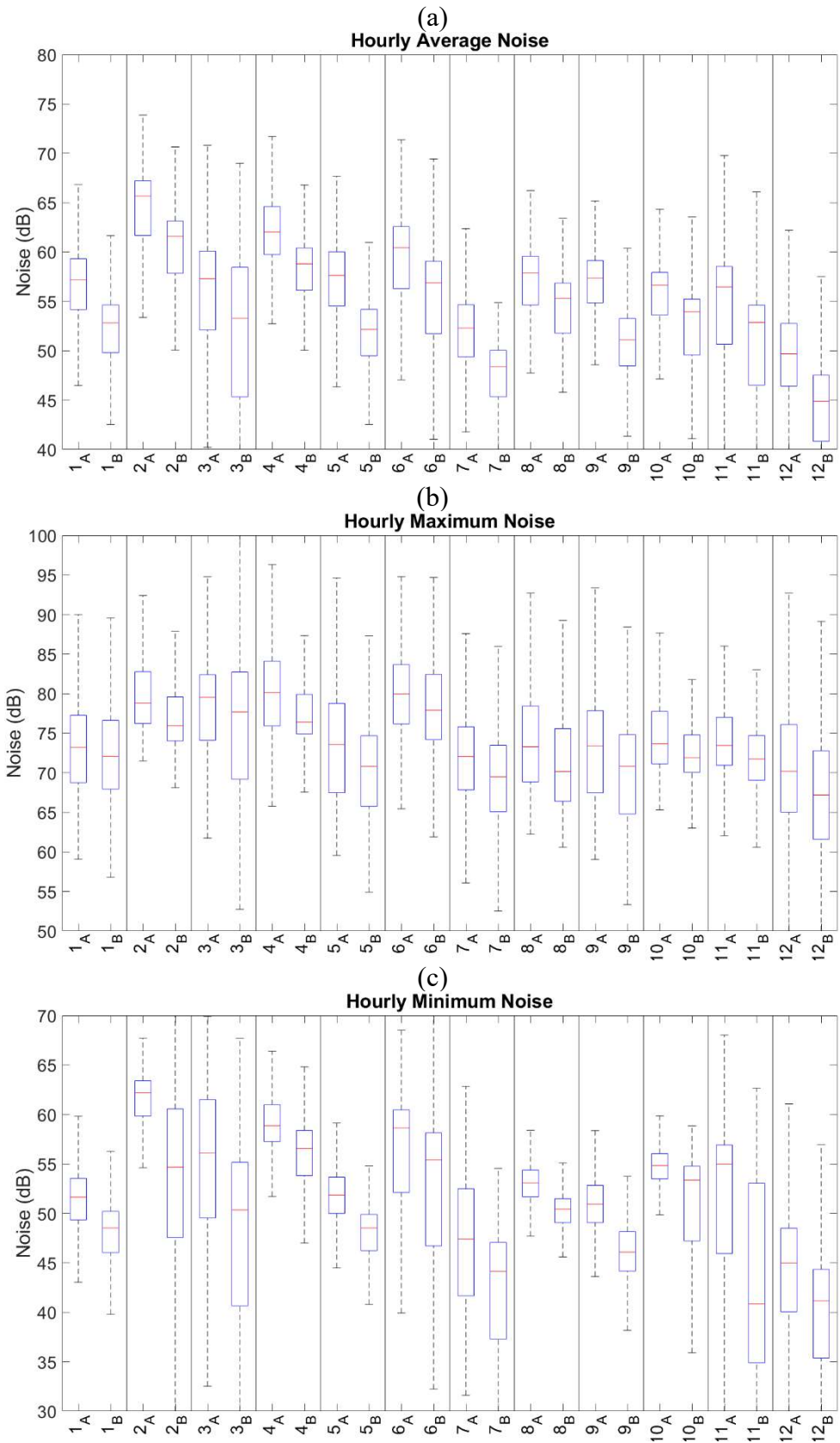

**Figure 3.** Boxplot of hourly (a) average, (b) maximum, and (c) minimum noise recorded pre lockdown (till 24th March 2020) and during lockdown period for each of the twelve noise monitoring stations.

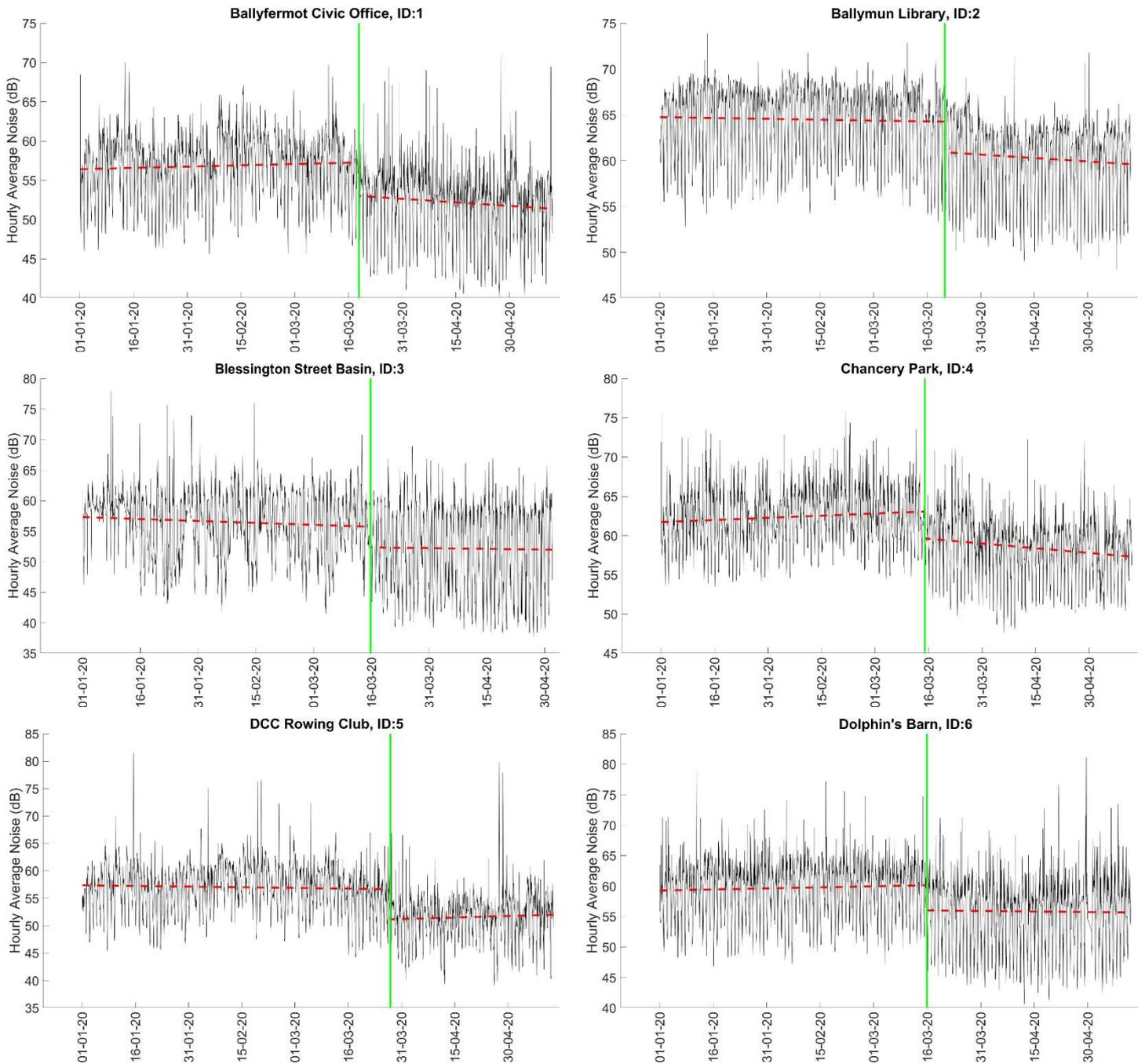

**Figure 4.** Hourly average noise time series plot (shown in black) and the identified change point for each noise monitoring station. The vertical green line indicates the day a change is noted, which the red dashed line denotes the linear trend of noise change over time.

(**Figure 4.** continued…)

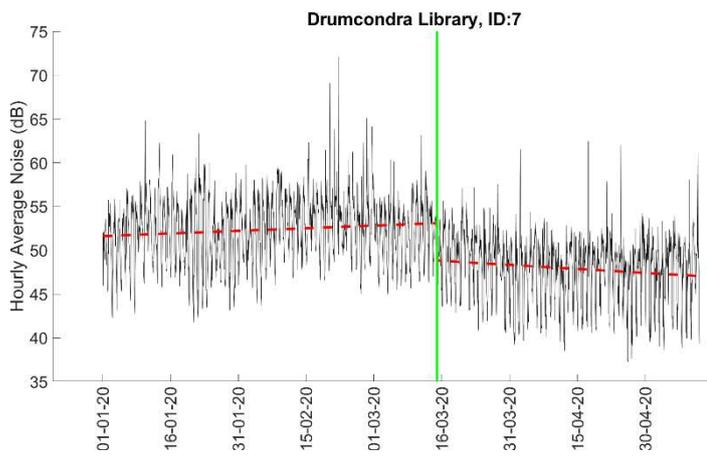
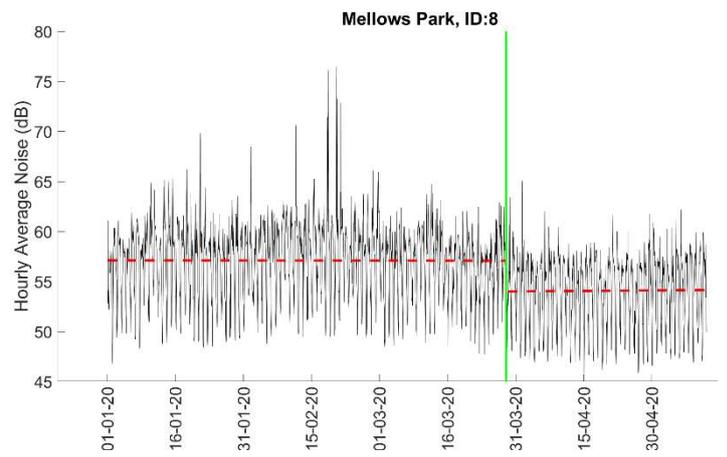
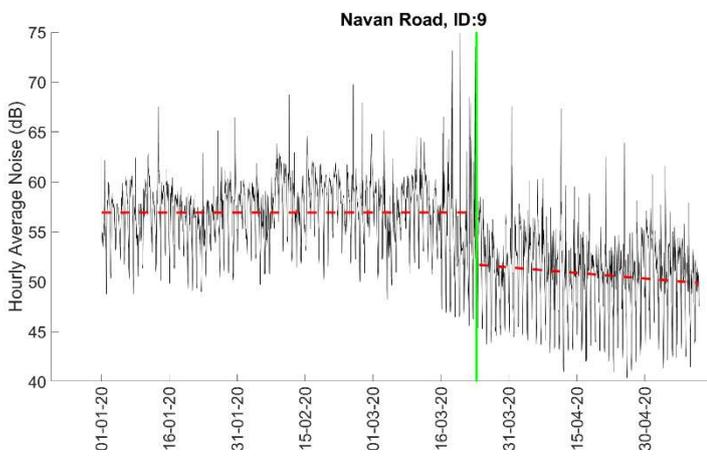
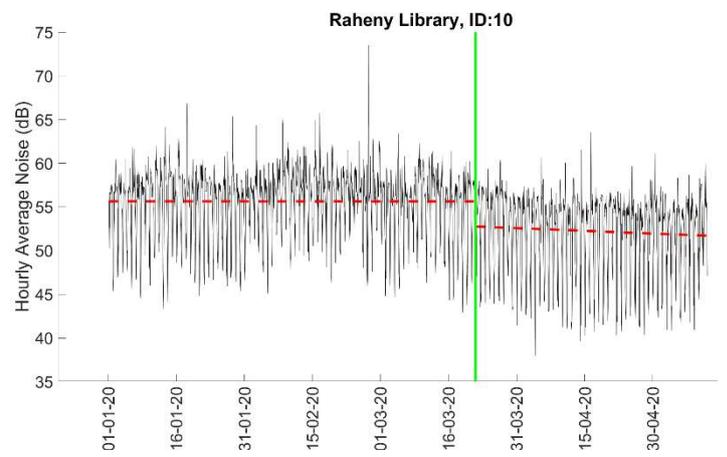
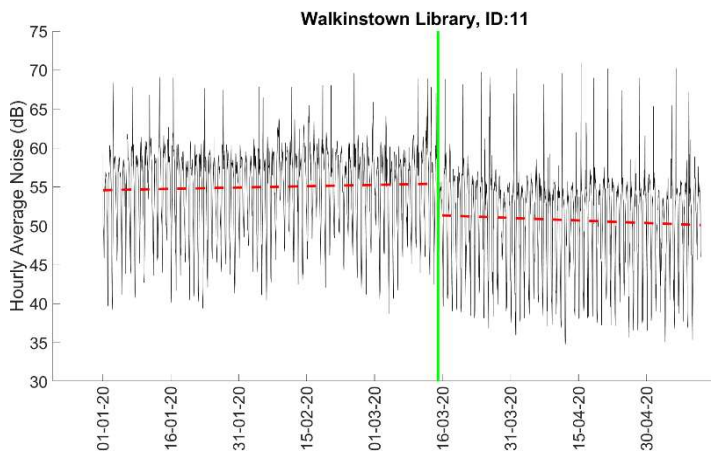
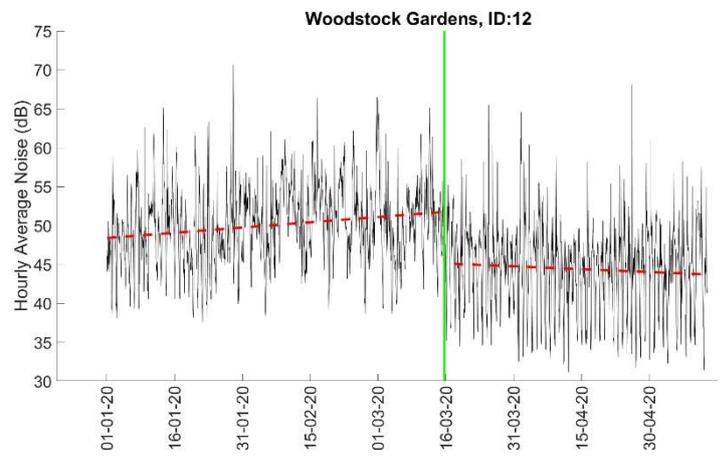

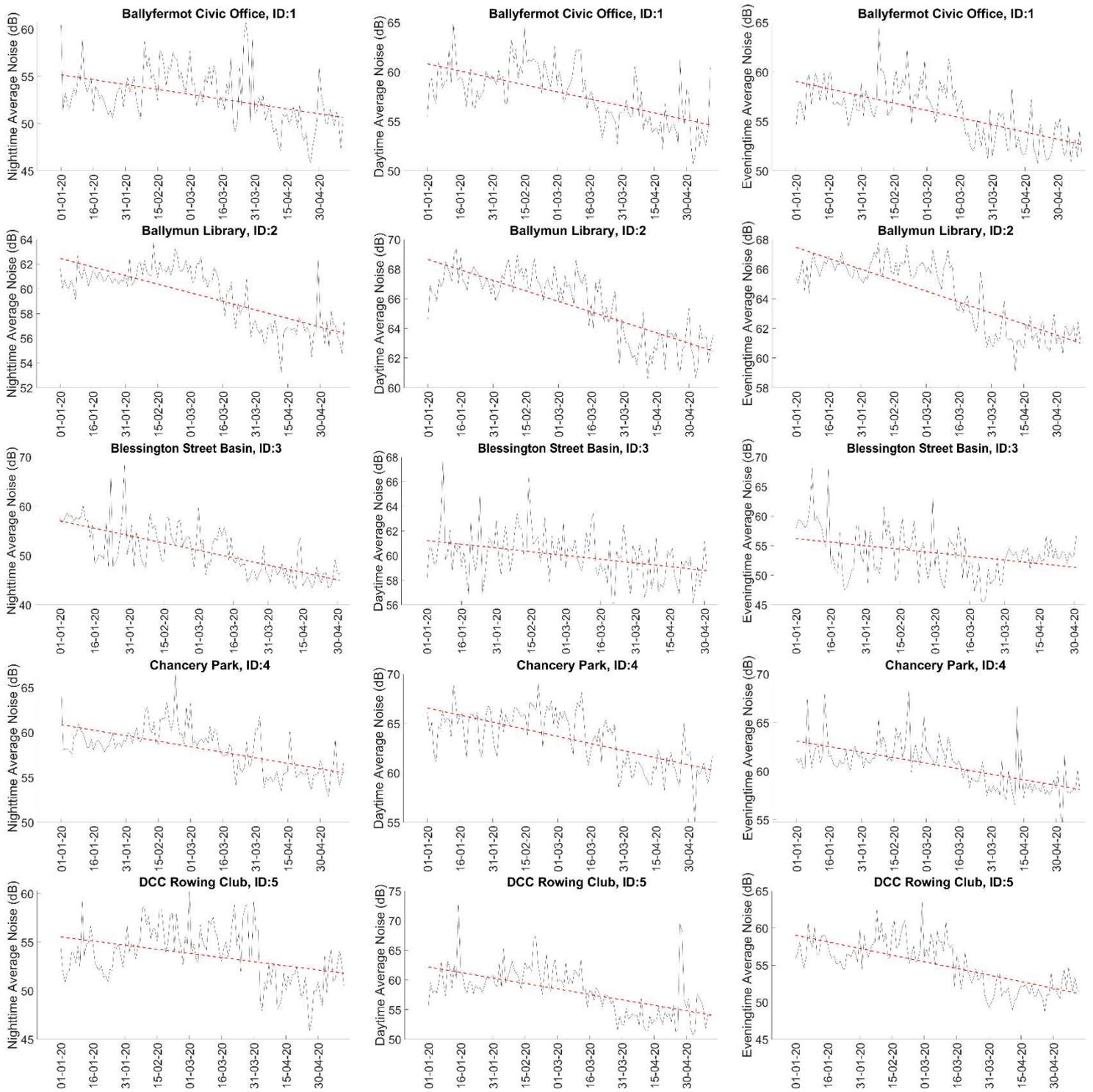

**Figure 5.** Time series plot of average noise during night time (11 PM to 7 AM), daytime (7 AM to 7 PM), and evening time (7 PM to 11 PM) for each noise monitoring station along with the linear trend.

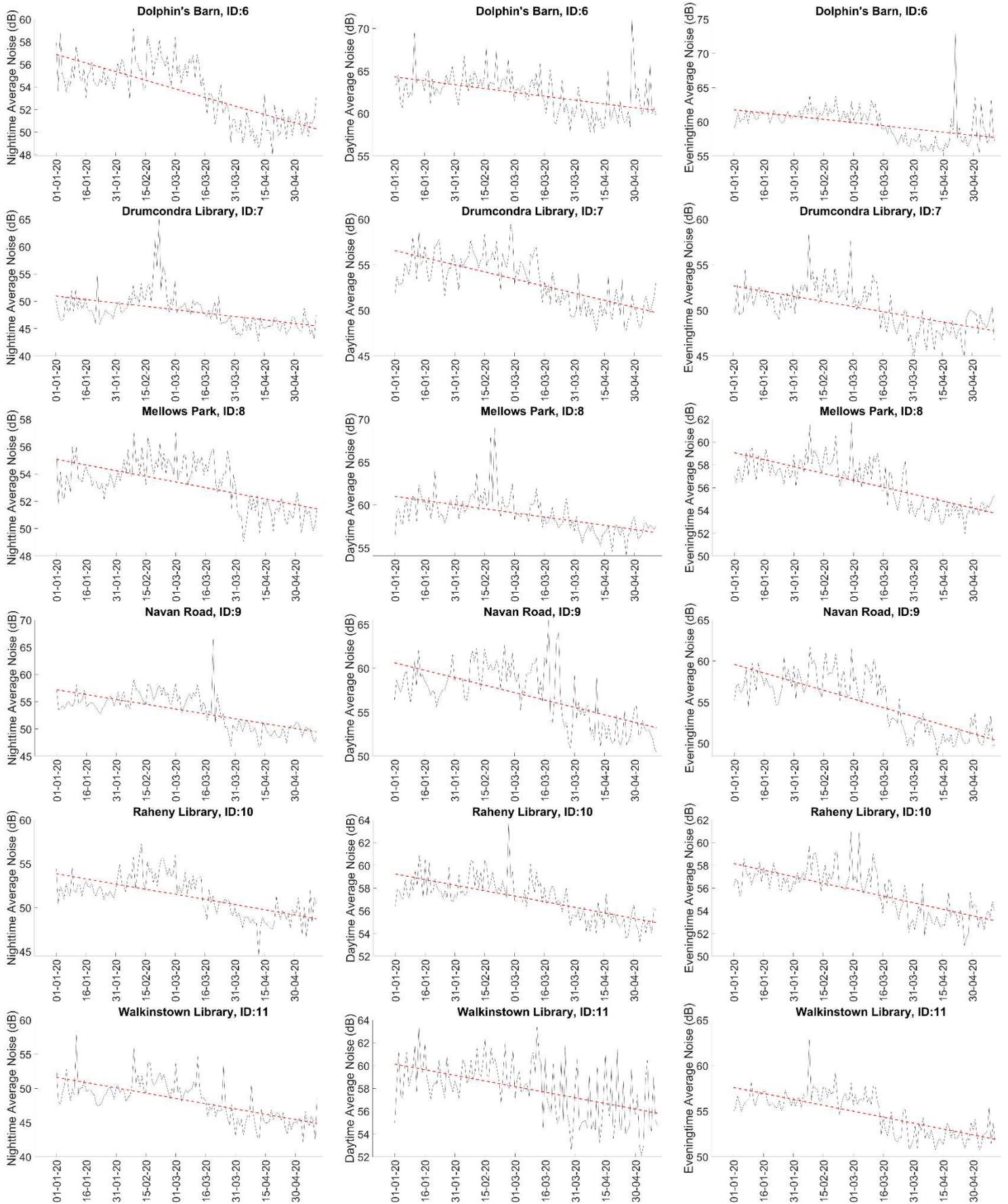

(**Figure 5.** continued…)

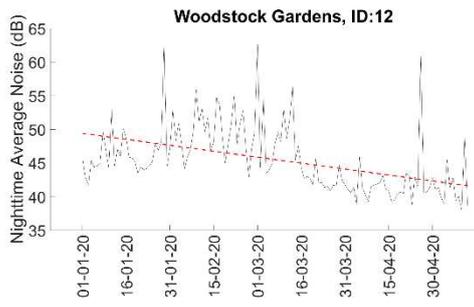 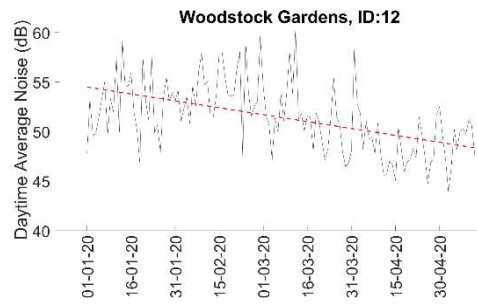 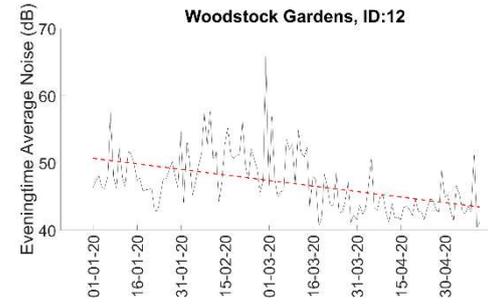

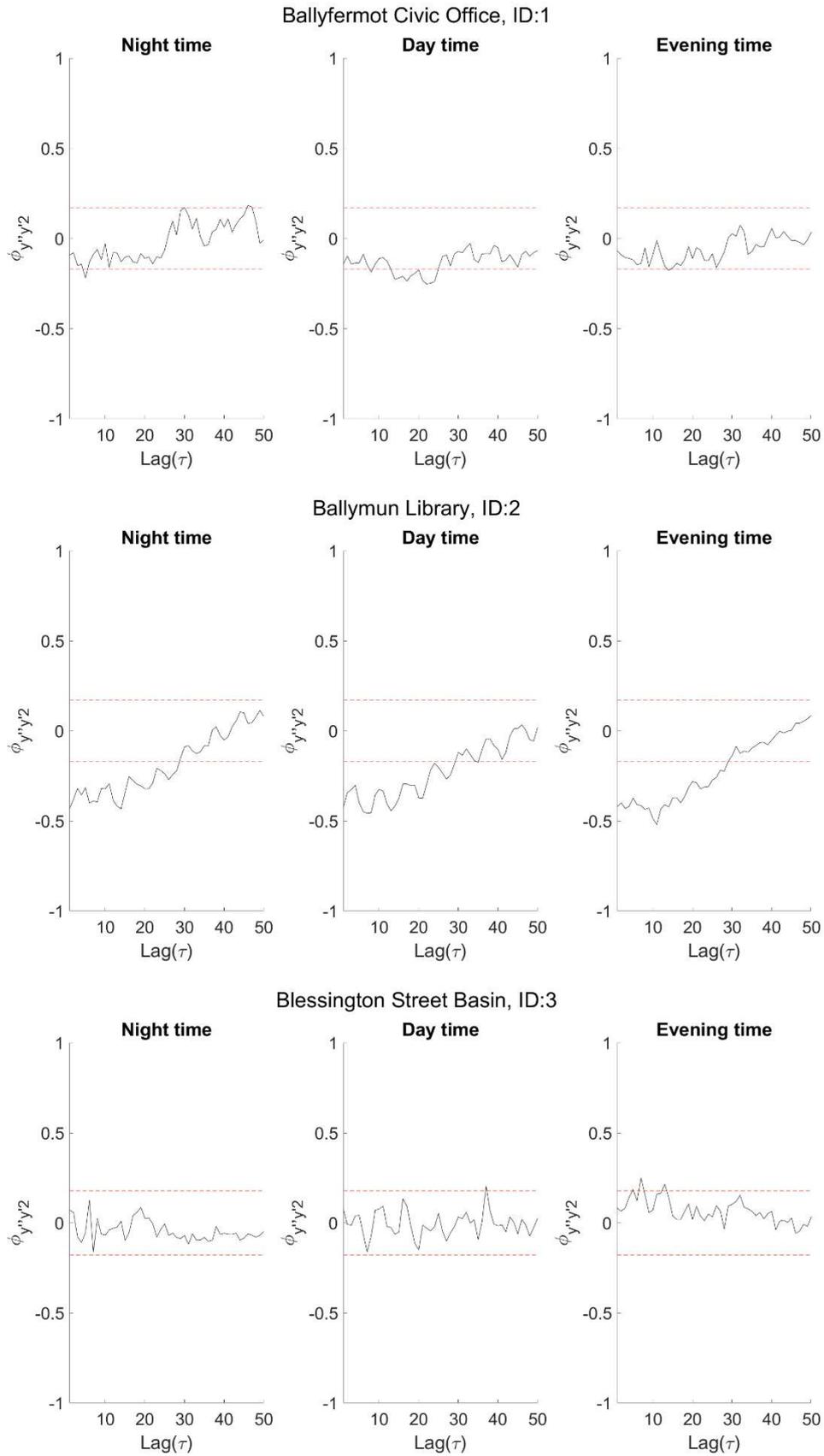

**Figure 6.** Cross-correlation function $\phi_{y',y'^2}(\tau)$ of night time, day time and evening time noise pollution data for each of the 12 noise monitoring stations corresponding to lag $\tau$.

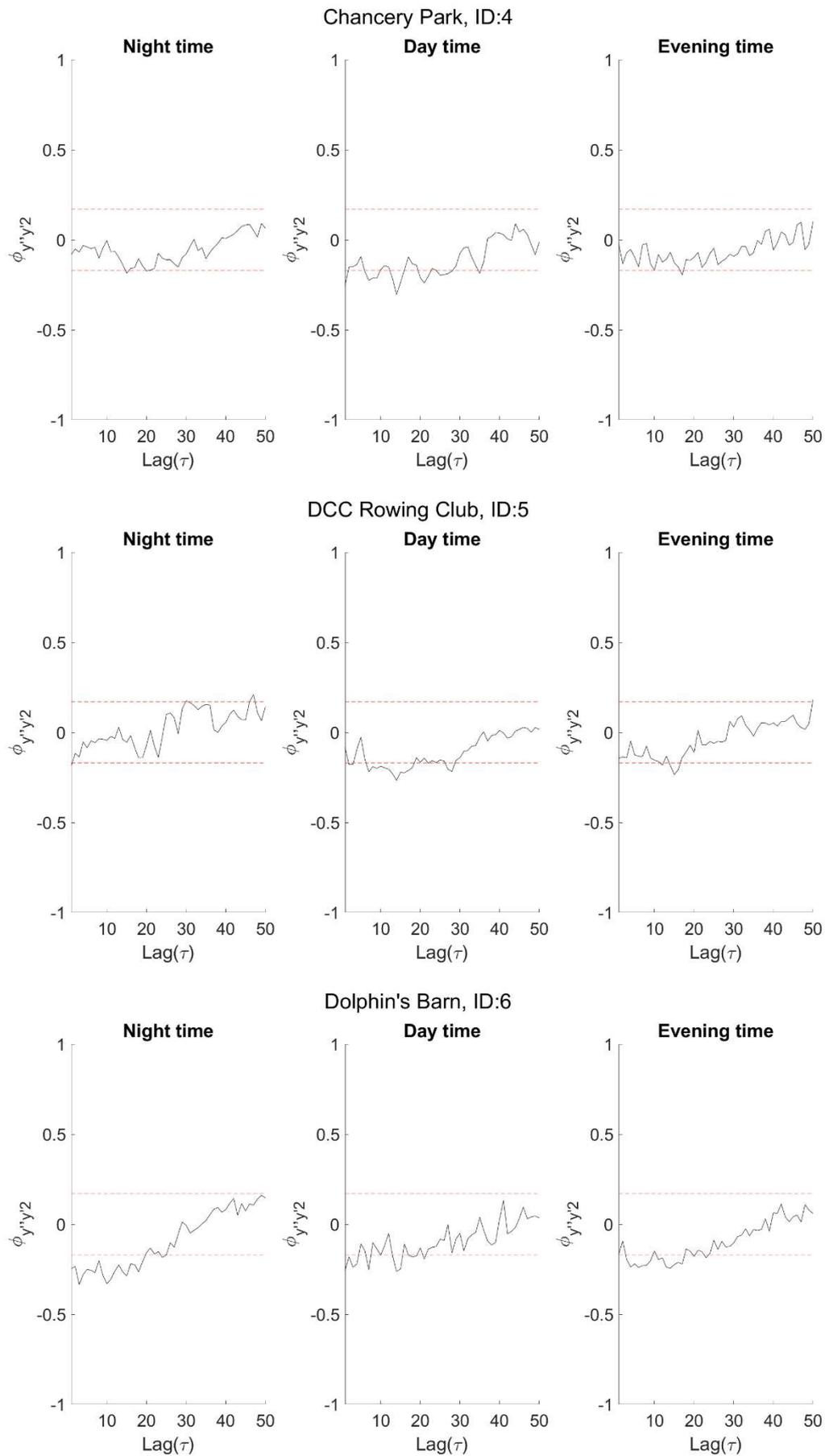

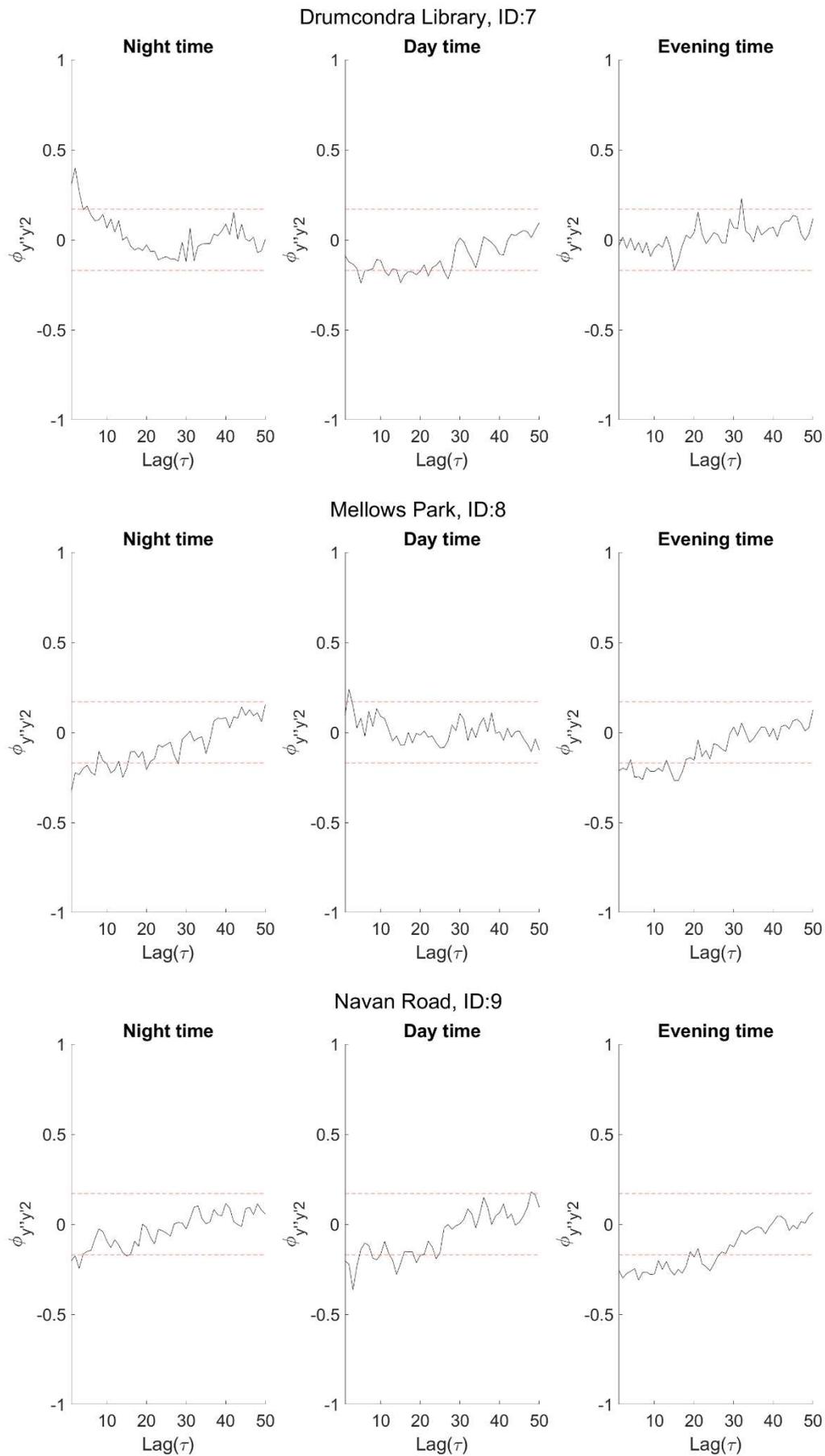

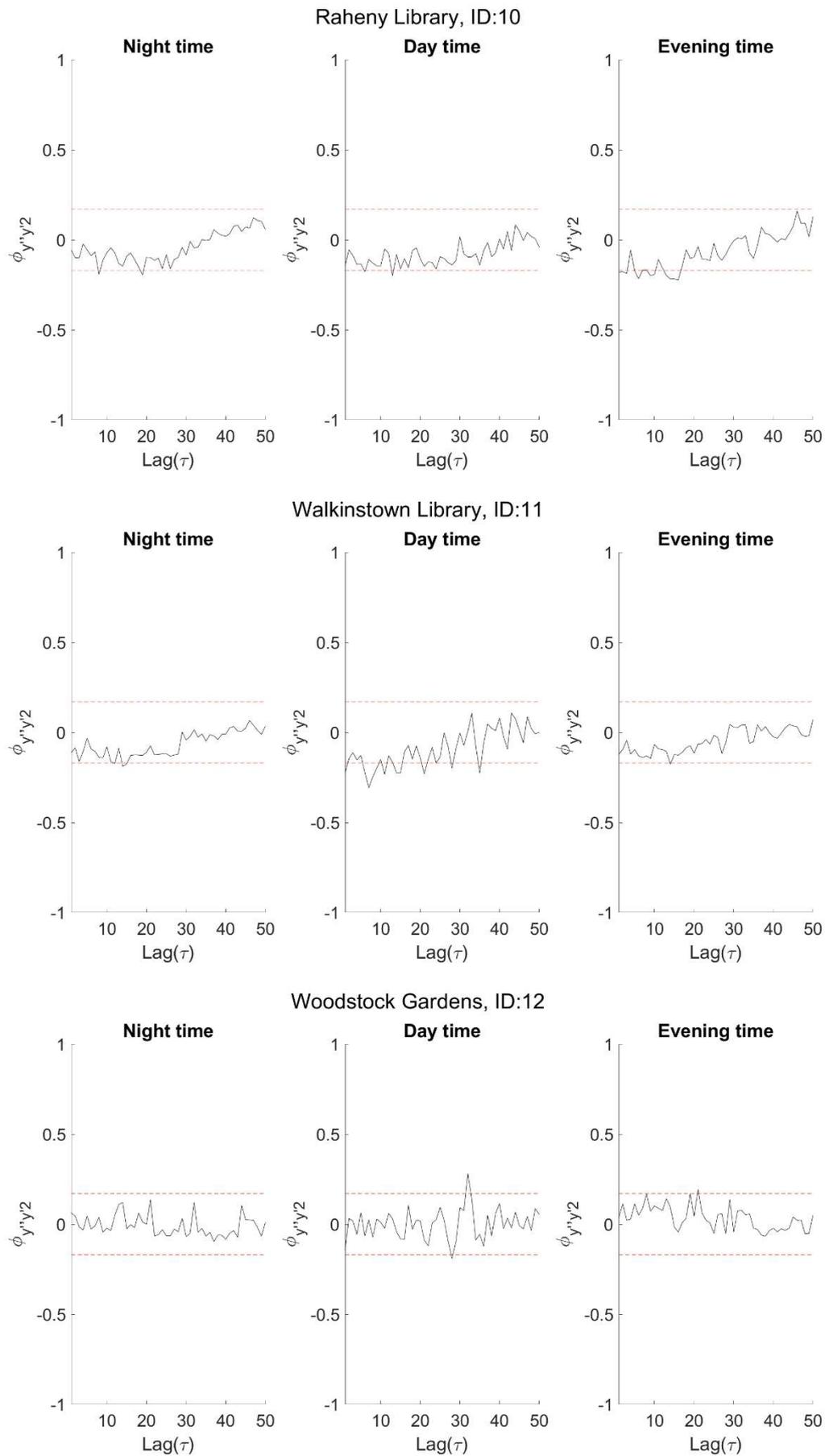

(a)

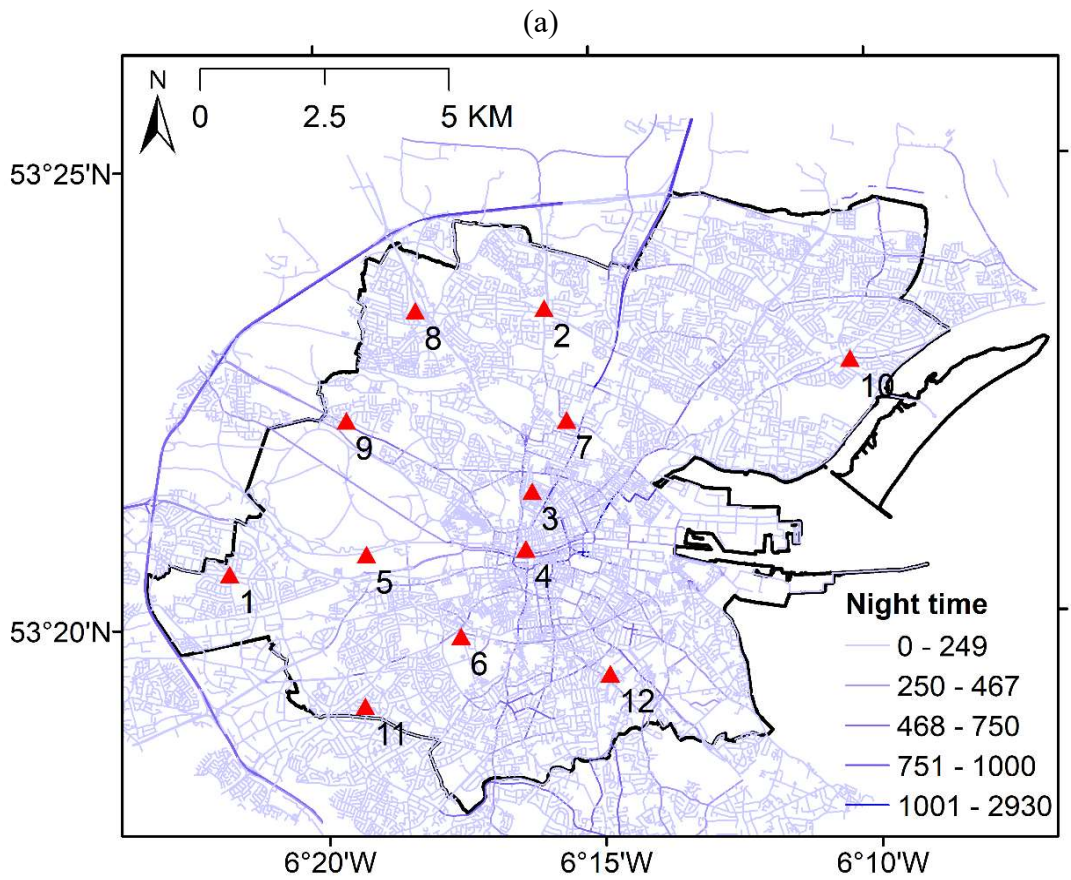

(b)

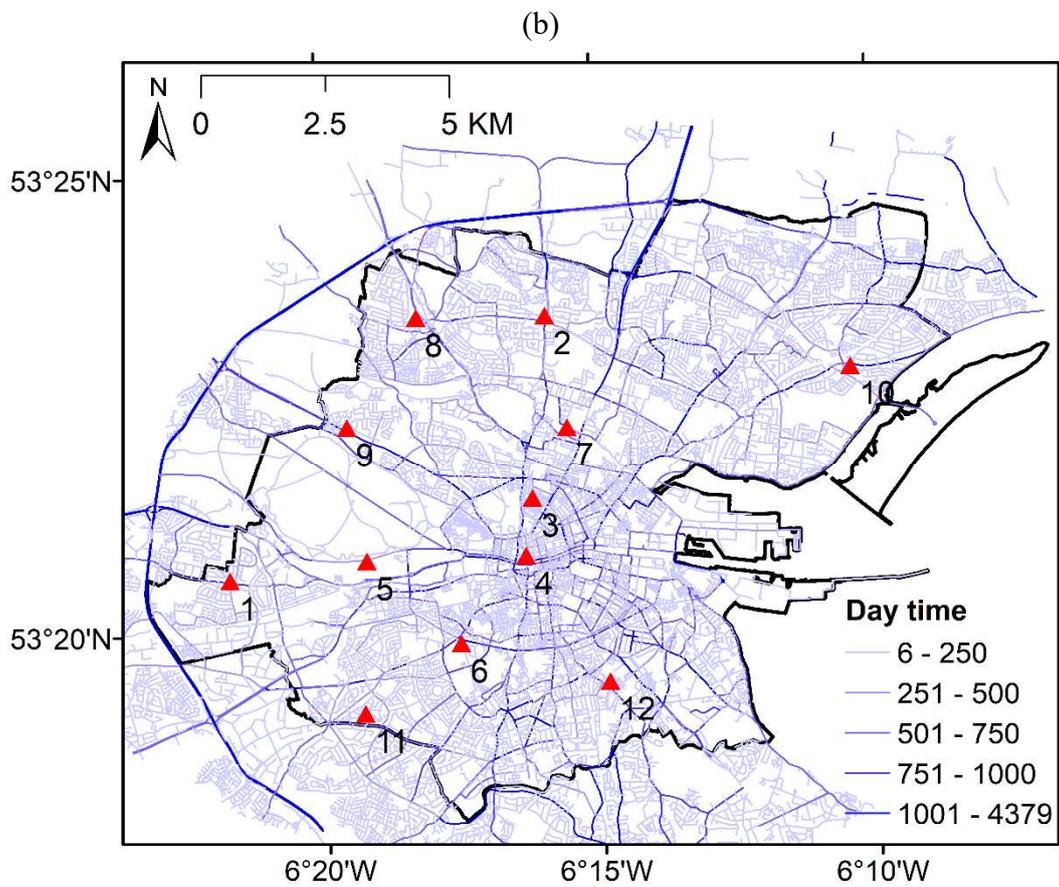

**Figure 7.** Average Traffic in Dublin during a) Night time, b) Day time and c) Evening time.

(**Figure 7.** continued…)

(c)

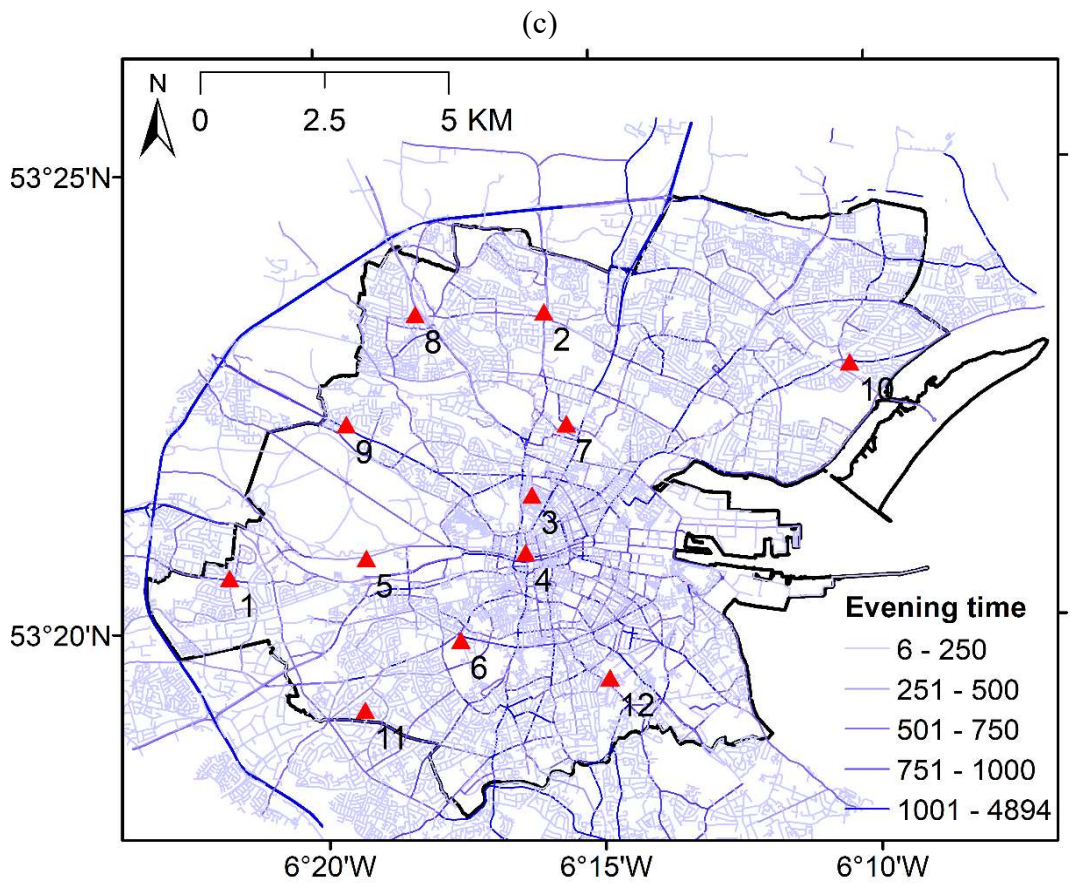

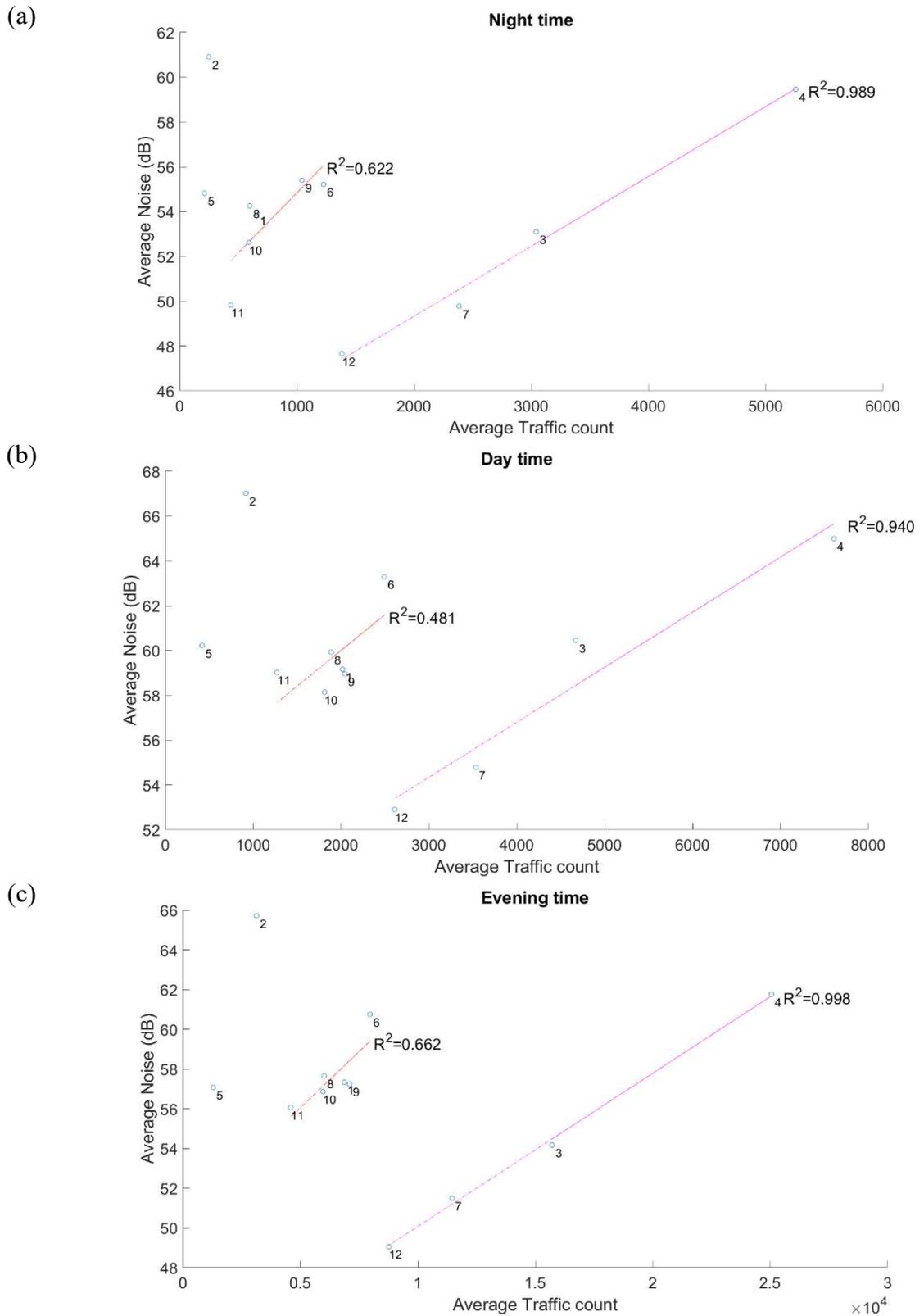

**Figure 8.** Scatter plot between average traffic count within a 500 meters radius and average noise pollution before lockdown period for each noise monitoring stations in Dublin corresponding to (a) night, (b) day, and (c) evening time period.

**Table 1.** Location details of twelve noise monitoring stations in Dublin, Ireland.

| Station ID | Name | Latitude (°N) | Longitude (°W) | Change date | No. of schools within 500 meter radius |
|---|---|---|---|---|---|
| 1 | Ballyfermot Civic Office | 53.343 | 6.362 | 19-Mar-2020 | 8 |
| 2 | Ballymun Library | 53.390 | 6.265 | 21-Mar-2020 | 5 |
| 3 | Blessington Street Basin | 53.357 | 6.270 | 16-Mar-2020 | 5 |
| 4 | Chancery Park | 53.347 | 6.272 | 15-Mar-2020 | 7 |
| 5 | DCC Rowing Club | 53.346 | 6.320 | 15-Mar-2020 | 2 |
| 6 | Dolphin's Barn | 53.331 | 6.292 | 16-Mar-2020 | 8 |
| 7 | Drumcondra Library | 53.370 | 6.259 | 15-Mar-2020 | 9 |
| 8 | Mellows Park | 53.391 | 6.304 | 29-Mar-2020 | 7 |
| 9 | Navan Road | 53.371 | 6.326 | 25-Mar-2020 | 0 |
| 10 | Raheny Library | 53.380 | 6.173 | 22-Mar-2020 | 5 |
| 11 | Walkinstown Library | 53.319 | 6.322 | 15-Mar-2020 | 0 |
| 12 | Woodstock Gardens | 53.324 | 6.248 | 17-Mar-2020 | 7 |

**Table 2.** Slope of changes in noise from 1st January to 11th May 2020 for hourly average and hourly maximum noise. The slope values denoted in bold are statistically significant at 5% significance level in two-tain hypothesis test. The p-values in the hypothesis test are shown in brackets. Negative value of slope indicates a decrease in noise pollution in time, while positive value indicate increase in noise pollution over time.

| Station ID | Average noise | | | Maximum noise | | | Minimum noise | | |
|---|---|---|---|---|---|---|---|---|---|
| | Night | Day | Evening | Night | Day | Evening | Night | Day | Evening |
| 1 | **-0.034** | **-0.047** | **-0.049** | *0.039* | -0.013 | -0.003 | **-0.083** | **-0.034** | **-0.035** |
| | (0.00) | (0.00) | (0.00) | (0.01) | (0.22) | (0.85) | (0.00) | (0.00) | (0.00) |
| 2 | **-0.046** | **-0.047** | **-0.049** | -0.004 | 0.008 | -0.021 | **-0.158** | **-0.082** | **-0.119** |
| | (0.00) | (0.00) | (0.00) | (0.69) | (0.35) | (0.07) | (0.00) | (0.00) | (0.00) |
| 3 | **-0.100** | **-0.020** | **-0.040** | -0.017 | 0.000 | *0.046* | **-0.141** | **-0.025** | **-0.116** |
| | (0.00) | (0.00) | (0.00) | (0.19) | (1.00) | (0.00) | (0.00) | (0.03) | (0.00) |
| 4 | **-0.041** | **-0.048** | **-0.038** | **-0.030** | **-0.059** | **-0.033** | **-0.067** | **-0.032** | **-0.030** |
| | (0.00) | (0.00) | (0.00) | (0.01) | (0.00) | (0.02) | (0.00) | (0.00) | (0.00) |
| 5 | **-0.028** | **-0.061** | **-0.060** | *0.037* | **-0.048** | **-0.044** | **-0.082** | **-0.046** | **-0.053** |
| | (0.00) | (0.00) | (0.00) | (0.00) | (0.00) | (0.00) | (0.00) | (0.00) | (0.00) |
| 6 | **-0.051** | **-0.030** | **-0.031** | **-0.033** | -0.022 | -0.007 | **-0.086** | **-0.048** | **-0.076** |
| | (0.00) | (0.00) | (0.00) | (0.00) | (0.13) | (0.43) | (0.00) | (0.00) | (0.00) |
| 7 | **-0.042** | **-0.052** | **-0.038** | **-0.035** | **-0.027** | 0.007 | **-0.051** | **-0.049** | **-0.054** |
| | (0.00) | (0.00) | (0.00) | (0.00) | (0.00) | (0.42) | (0.00) | (0.00) | (0.00) |
| 8 | **-0.028** | **-0.032** | **-0.040** | -0.011 | **-0.032** | **-0.032** | **-0.048** | **-0.029** | **-0.032** |
| | (0.00) | (0.00) | (0.00) | (0.31) | (0.00) | (0.00) | (0.00) | (0.00) | (0.00) |
| 9 | **-0.059** | **-0.056** | **-0.070** | 0.005 | **-0.078** | **-0.053** | **-0.138** | **-0.052** | **-0.063** |
| | (0.00) | (0.00) | (0.00) | (0.69) | (0.00) | (0.00) | (0.00) | (0.00) | (0.00) |
| 10 | **-0.039** | **-0.033** | **-0.039** | 0.005 | **-0.050** | -0.011 | **-0.067** | **-0.012** | **-0.053** |
| | (0.00) | (0.00) | (0.00) | (0.60) | (0.00) | (0.32) | (0.00) | (0.00) | (0.00) |
| 11 | **-0.051** | **-0.033** | **-0.043** | 0.001 | 0.002 | **-0.029** | **-0.094** | **-0.094** | **-0.106** |
| | (0.00) | (0.00) | (0.00) | (0.95) | (0.90) | (0.01) | (0.00) | (0.00) | (0.00) |
| 12 | **-0.059** | **-0.046** | **-0.055** | -0.010 | -0.017 | -0.011 | **-0.102** | **-0.018** | **-0.071** |
| | (0.00) | (0.00) | (0.00) | (0.54) | (0.19) | (0.43) | (0.00) | (0.01) | (0.00) |

**Table 3.** Average hourly traffic count within 500 meters radius surrounding each station in the night, day, and evening time obtained from DCC SCATS.

| Station ID | Night | Day | Evening |
|---|---|---|---|
| 1 | 658.5 | 2018.1 | 6877.6 |
| 2 | 249.9 | 917.9 | 3138.8 |
| 3 | 3042.4 | 4670.2 | 15721.5 |
| 4 | 5256.1 | 7608.7 | 25063.0 |
| 5 | 212.6 | 420.2 | 1293.5 |
| 6 | 1228.9 | 2492.7 | 7969.0 |
| 7 | 2385.2 | 3534.5 | 11461.5 |
| 8 | 599.9 | 1889.5 | 6016.5 |
| 9 | 1043.5 | 2043.7 | 7100.3 |
| 10 | 594.0 | 1813.9 | 5971.3 |
| 11 | 437.5 | 1272.4 | 4597.9 |
| 12 | 1385.4 | 2611.1 | 8776.3 |

**Table 4.** Number and percentage of time the recorded hourly average noise exceeded the threshold 55 dB. Numbers in bold denote lowest noise in every station.

| Station ID | Total number | | Percentage | |
|---|---|---|---|---|
| | Pre lockdown | During lockdown | Pre lockdown | During lockdown |
| 1 | 1393 | 247 | 69.13 | **21.42** |
| 2 | 1987 | 963 | 98.61 | **83.52** |
| 3 | 1244 | 420 | 61.74 | **46.00** |
| 4 | 1962 | 947 | 97.37 | **82.13** |
| 5 | 1467 | 203 | 72.80 | **17.29** |
| 6 | 1637 | 700 | 81.24 | **62.00** |
| 7 | 432 | 8 | 21.44 | **0.69** |
| 8 | 1466 | 616 | 72.75 | **53.43** |
| 9 | 1491 | 121 | 74.00 | **10.49** |
| 10 | 1387 | 340 | 69.11 | **29.49** |
| 11 | 1207 | 220 | 59.90 | **19.08** |
| 12 | 265 | 29 | 13.15 | **2.47** |